# R.E. Axiomatization of Conditional Independence

Pavel Naumov

Department of Mathematics
and Computer Science
McDaniel College
Westminster, Maryland, USA
pnaumov@mcdaniel.edu

Brittany Nicholls

Department of Mathematics
and Computer Science
McDaniel College
Westminster, Maryland, USA
brn002@mcdaniel.edu

## ABSTRACT

The paper investigates properties of the conditional independence relation between pieces of information. This relation is also known in the database theory as embedded multivalued dependency. In 1980, Parker and Parsaye-Ghomi established that the properties of this relation can not be described by a finite system of inference rules. In 1995, Herrmann proved that the propositional theory of this relation is undecidable. The main result of this paper is a complete recursively enumerable axiomatization of this theory.

## 1. INTRODUCTION

In this paper, we study the properties of interdependencies between pieces of information. We call these pieces *secrets* to emphasize the fact that they might be unknown to some parties. For example, if secret $a$ is the area of a triangle and secret $p$ is the perimeter of the same triangle, then there is an interdependence between these secrets in the sense that not every value of secret $a$ is compatible with every value of secret $p$. If there is no interdependence between two secrets, then we say that the two secrets are *independent*. In other words, secrets $a$ and $b$ are independent if each possible value of secret $a$ is compatible with each possible value of secret $b$. We denote this relation between two secrets by $a \parallel b$. This relation was introduced by Sutherland [18] and is sometimes referred to as *nondeducibility*. Halpern and O'Neill [6] proposed a closely related notion called $f$-secrecy. Donders, More, and Naumov described properties of a multi-argument variation $a_1 \parallel a_2 \parallel \cdots \parallel a_n$ of the same relation under the assumption that the secrets are generated over an undirected graph [12], a directed acyclic graph [2], or a hypergraph [11] with a fixed topology.

Independence relation can be generalized to relate two sets of secrets. If $A$ and $B$ are two such sets, then $A \parallel B$ means that any consistent combination of values of secrets in set $A$ is compatible with any consistent combination of values of secrets in set $B$. Note that "consistent combination" is an important condition here since some interdependence may exist between secrets in set $A$ even while the entire set of secrets $A$ is independent from the secrets in set $B$. A sound and complete axiomatization of this relation between sets of secrets was given by More and Naumov [10]:

1. *Empty Set*: $\varnothing \parallel A$,

2. *Monotonicity*: $A, B \parallel C \to A \parallel C$,

3. *Symmetry*: $A \parallel B \to B \parallel A$,

4. *Exchange*: $A, B \parallel C \to (A \parallel B \to A \parallel B, C)$,

where here and everywhere below by $A, B$ we mean the union of the sets $A$ and $B$. The same axioms were shown by Geiger, Paz, and Pearl [3] to provide a complete axiomatization of the independence relation between sets of random variables in probability theory. More recently, the same system was shown to be sound an complete with respect to concurrency [14] and game [15] semantics.

Suppose now that $a$, $b$, $c$, and $d$ are four secrets with integer values such that $a+b+c+d \equiv 0 \pmod 2$. Note that $a \parallel b$ is true since every possible value of $a$ is consistent with any possible value of $b$. At the same time, if values of $c$ and $d$ are fixed, then not every possible value of secret $a$ is compatible with every possible value of secret $b$. We will say that secrets $a$ and $b$ are not independent conditionally on $c, d$ and denote this by $\neg(a \parallel_{c,d} b)$. On the other hand, if only value of $c$ is fixed, then any value of $a$ is still consistent with any value of $b$. We write this as $a \parallel_c b$. In general, conditional independence relation $A \parallel_C B$ can be defined between any three disjoint sets of secrets. This relation, which is also known in the database theory as *embedded multivalued dependency*, has many non-trivial properties. For example, later we will show soundness of the following principles:

$$A \parallel_C B \wedge A \parallel_{B,C} D \to A \parallel_C B, D,$$

$$A, B \parallel_C D \to A \parallel_{B,C} D,$$

$$B \parallel_A C \wedge E \parallel_B D \wedge D \parallel_C F \wedge E \parallel_D F \wedge A \parallel_E F \to E \parallel_A F.$$

Parker and Parsaye-Ghomi [16] have shown that this relation can not be described by a finite system of inference rules. Herrmann [7, 8] proved the undecidability of the propositional theory of this relation. Lang, Liberatore, and Marquis [9] studied complexity of conditional independence between sets of propositional variables. Studený [17] has shown that the related conditional independence in probability theory has no complete finite characterization. More recently, Grädel and Väänänen discussed (incomplete) logical systems describing properties of the conditional independence in propositional and first order languages [4] and suggested model checking game semantics for these systems [5].

The main result of this paper is a complete infinite recursively enumerable axiomatization of the propositional theory of the relation $A \parallel_C B$. This work builds on the techniques from our previous TARK paper [13], where we gave a complete axiomatization of a different ternary knowledge relation. The "diagram" notion used in the current paper is a generalization of the "diamond" notations from the previous paper.

## 2. SYNTAX AND SEMANTICS

We assume a fixed alphabet of "secret" variables: $a, b, \ldots$.

DEFINITION 1. *By the set of formulas $\Phi$ we mean the minimal set of formulas such that*

1. $\bot \in \Phi$,

2. $A \parallel_C B \in \Phi$ *for each pairwise disjoint sets of secret variables $A$, $B$, and $C$,*

3. $\varphi_1 \to \varphi_2 \in \Phi$ *if $\varphi_1, \varphi_2 \in \Phi$.*

As usual, all other boolean connectives are assumed to be defined through the implication and the constant false.

DEFINITION 2. *A protocol is a pair $\mathcal{P} = \langle V, R \rangle$, where,*

1. *for any secret variable $a$, set $V(a)$ is an arbitrary set of "values" of secret $a$,*

2. *$R$ is a set of functions $r$ on secret variables such that $r(a) \in V(a)$ for any secret variable $a$. Elements of $R$ will be called "runs" of the protocol.*

For any set of secret variables $A$ and any runs $r_1$ and $r_2$, we write $r_1 \equiv_A r_2$ if $r_1(a) = r_2(a)$ for any $a \in A$. The next definition is the core definition of this paper. Item 3 below formally defines conditional independence relation between sets of secrets.

DEFINITION 3. *For any protocol $\mathcal{P} = \langle V, R \rangle$ and any formula $\varphi \in \Phi$, we define the binary relation $\mathcal{P} \vDash \varphi$ as follows:*

1. $\mathcal{P} \nvDash \bot$,

2. $\mathcal{P} \vDash A \parallel_C B$ *if and only if, for any $r_1, r_2 \in R$, such that $r_1 \equiv_C r_2$, there is $r \in R$ such that $r_1 \equiv_{A,C} r \equiv_{B,C} r_2$.*

3. $\mathcal{P} \vDash \varphi \to \psi$ *if and only if $\mathcal{P} \nvDash \varphi$ or $\mathcal{P} \vDash \psi$.*

## 3. GRAPH NOTATIONS

In this paper we deal with graphs that might have directed as well as undirected edges. An example $G$ of such graph is depicted in Figure 1. We use word "path" for any sequences of adjacent vertices without taking into account the directions of edges. For example, sequence of vertices $v_1, v_2, v_3$ is a path in graph $G$. The graphs that we consider are "labeled". By that we mean that each edge of the graph is labeled with a *set* of secret variables. If vertices $u$ and $w$ of the graph are connected by a path such that each edge of the path is labeled with a set containing label $x$, then we write $u \sim_x w$. For example, $v_1 \sim_a v_3$ in graph $G$. We allow paths that consist of just a single vertex. This assumption implies that relation $u \sim_x w$ is an equivalence relation on graphs for any fixed label $x$.

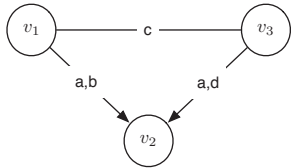

**Figure 1: Graph $G$**

For any *set* of labels $X$, we write $u \sim_X w$ if $u \sim_x w$ for each $x \in X$. For example, $v_1 \sim_{a,c} v_3$ in graph $G$. Note that $a$-path and $c$-path from $v_1$ to $v_3$ are not the same. Relation $u \sim_X w$ is also an equivalence relation on vertices for any fixed set of secret variables $X$. Sometimes we draw only a fragment of a graph. To show that vertices $u$ and $w$ are in relation $u \sim_X w$ on the whole graph, we connect vertices $u$ and $w$ in our partial drawing by a *double line* labeled with set $X$. For example,

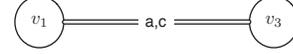

is a partial drawing of the graph $G$ from Figure 1.

## 4. DIAGRAMS

The description of the axiomatic system for conditional independence proposed in this paper is using the notion of a diagram. Informal drawing similar to our diagrams have been used before to visualize arguments about *specific* properties of conditional independence. See, for example, illustrations in Parker and Parsaye-Ghomi [16]. In this work, however, we give such drawings a precise mathematical definition and show, through the proof of completeness theorem, that *all* properties of conditional independence can be observed by analyzing the diagrams.

A diagram is a labeled graph with a special structure. For each diagram $\Delta$ there is a set of formulas $[\Delta]$ that, informally, is used to "construct" the diagram. Formally, the diagrams and the corresponding sets of formulas are defined below.

DEFINITION 4. *For any set of secret variables $Q$, the set of diagram $Diag(Q)$ is the minimal set such that*

1. *it contains the "basic" diagram $\Delta_0$ consisting of two vertices, called $v^+$ and $v^-$, and an undirected edge between $v^+$ and $v^-$ labeled with $Q$:*

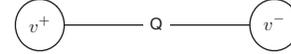

*By definition, set $[\Delta_0]$ is empty.*

2. *For any pair-wise disjoint sets $A$, $B$, and $C$, and any two vertices $u$ and $v$ of a diagram $\Delta \in Diag(Q)$, such that $u \sim_C v$, there is a diagram $\Delta' \in Diag(Q)$:*

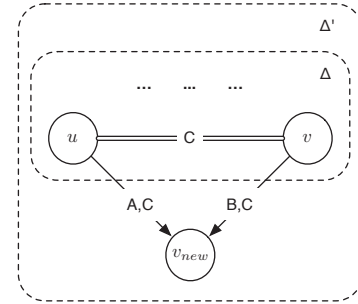

*such that*

(a) *Diagram $\Delta'$, in addition to all vertices of the diagram $\Delta$, contains a new vertex $v_{new}$,*

(b) *Diagram $\Delta'$, in addition to all edges of the diagram $\Delta$, contains two new directed edges $(u, v_{new})$ and $(v, v_{new})$ labeled by sets $A \cup C$ and $B \cup C$ respectively.*

(c) $[\Delta'] = [\Delta] \cup \{A \parallel_C B\}$.

If diagrams $\Delta$ and $\Delta'$ are related as described above, then we say that diagram $\Delta'$ is an extension of the diagram $\Delta$. The same diagram $\Delta$ has multiple extensions. The unique vertices $v^+$ and $v^-$ from which construction of a diagram $\Delta$ was started will be referred to as $v_\Delta^+$ and $v_\Delta^-$. Note that if $\Delta \in Diag(Q)$, then $v_\Delta^+ \sim_Q v_\Delta^-$.

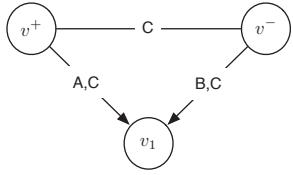

**Figure 2: Diagram $\Delta_1 \in Diag(C)$**

For example, diagram $\Delta_1$ in Figure 2 is obtained from the basic diagram through a single extension using sets $A$, $B$, and $C$. Thus, $[\Delta_1] = \{A \parallel_C B\}$.

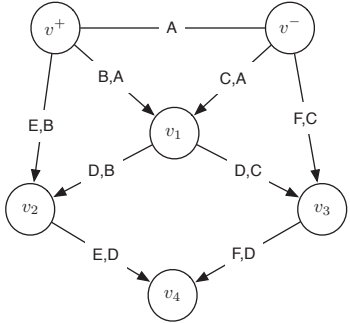

**Figure 3: Diagram $\Delta_2 \in Diag(A)$**

On the other hand, diagram $\Delta_2$ in Figure 3 can be constructed from the basic diagram by first adding vertex $v_1$, next $v_2$, next $v_3$, and finally $v_4$. Alternatively, the order can be $v_1$, $v_3$, $v_2$, and $v_4$. In either case, $[\Delta_2] = \{(B \parallel_A C), (E \parallel_B D), (D \parallel_C F), (E \parallel_D F)\}$. Note that vertex $v_4$ was added in spite of the lack of a direct edge from vertex $v_2$ to vertex $v_3$. For the diagram to extand to $v_4$ we only require $v_2 \sim_D v_3$.

DEFINITION 5. *Let*
$$t = (A_1, A_2, A_3; B_1, B_2, B_3; C_1, C_2, C_3; D)$$
*be a tuple of disjoint sets of labels. We say that diagram $\Delta \in Diag(C_1 \cup C_2 \cup C_3)$ renders tuple $t$ if diagram $\Delta$ contain vertices $w_1$ and $w_2$ such that*

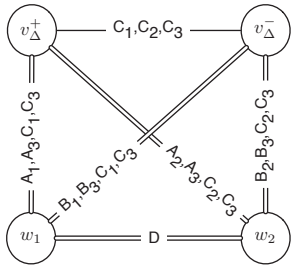

*or, in other words, $w_1 \sim_{A_1,A_3,C_1,C_3} v_\Delta^+$; $w_1 \sim_{B_1,B_3,C_1,C_3} v_\Delta^-$; $w_2 \sim_{A_2,A_3,C_2,C_3} v_\Delta^+$; $w_2 \sim_{B_2,B_3,C_2,C_3} v_\Delta^-$; $w_1 \sim_D w_2$.*

For example, Diagram $\Delta_1$, depicted in Figure 2, renders (with $w_1 = v_1$ and $w_2 = v^+$) tuple
$$(A, \varnothing, \varnothing; \varnothing, D, B; \varnothing, \varnothing, C; \varnothing)$$
for an arbitrary set of secrets $D$, because

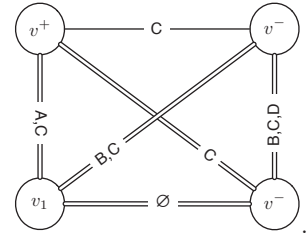

Here, of course, we use the fact that $v^- \sim_{B,C,D} v^-$ for each set of secrets $D$.

As another example, Diagram $\Delta_2$, depicted in Figure 3, renders (with $w_1 = v^+$ and $w_2 = v_4$) tuple
$$(\varnothing, \varnothing, E; \varnothing, F, \varnothing; A, \varnothing, \varnothing; \varnothing),$$
because

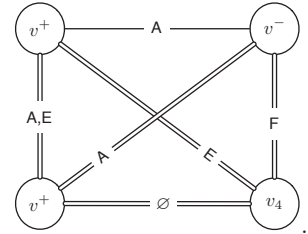

## 5. AXIOMS

In this section we introduce a logical system describing properties of conditional independence. The axioms of the system are:

1. Symmetry: $A \parallel_C B \to B \parallel_C A$,

2. Monotonicity: $A \parallel_C B, D \to A \parallel_C B$,

3. Diagram:
$$\wedge[\Delta] \to (A_1, B_1, C_1 \parallel_{A_3,B_3,C_3,D} A_2, B_2, C_2 \to$$
$$A_1, A_2, A_3 \parallel_{C_1,C_2,C_3} B_1, B_2, B_3),$$

if diagram $\Delta$ renders tuple
$$(A_1, A_2, A_3; B_1, B_2, B_3; C_1, C_2, C_3; D)$$
and $\wedge[\Delta]$ stands for conjunction of all formulas in $[\Delta]$.

We write $\vdash \varphi$ if formula $\varphi \in \Phi$ is provable from the above axioms and propositional tautologies in the language $\Phi$ using Modes Ponens inference rule. We write $X \vdash \varphi$ if formula $\varphi$ is provable in our logical system using an additional set of axioms $X$.

THEOREM 1. *The set of axioms of this logical system is recursively enumerable.*

PROOF. The statement of the theorem follows from recursive enumerability of diagrams, recursive enumerability of tuples, and decidability of "diagram renders tuple" relation. □

# 6. EXAMPLES

In this section we give several examples of formal proofs in our logical system. The soundness of the axioms will be shown in Section 7. We start with the three non-trivial properties of the conditional independence mentioned in the introduction.

PROPOSITION 1. $\vdash A \parallel_C B \wedge A \parallel_{B,C} D \to A \parallel_C B, D$.

PROOF. Consider diagram $\Delta_1$ depicted in Figure 2. As we have shown in Section 4, this diagram renders tuple $(A, \varnothing, \varnothing; \varnothing, D, B; \varnothing, \varnothing, C; \varnothing)$. Thus, by the Diagram axiom,

$$\vdash [\Delta_1] \to (A \parallel_{B,C} D \to A \parallel_C B, D).$$

Recall from Section 4 that $[\Delta_1] = \{A \parallel_C B\}$. Therefore, $\vdash A \parallel_C B \to (A \parallel_{B,C} D \to A \parallel_C B, D)$. □

PROPOSITION 2. $\vdash A, B \parallel_C D \to A \parallel_{B,C} D$.

PROOF. Consider basic diagram $\Delta_3$:

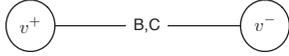

This diagram renders (with $w_1 = v^+$ and $w_2 = v^-$) tuple $(A, \varnothing, \varnothing; \varnothing, D, \varnothing; B, \varnothing, C; \varnothing)$ because

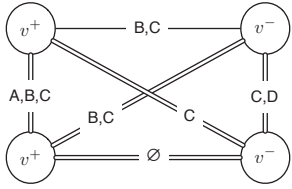

Hence, by the Diagram axiom,

$$\vdash \wedge[\Delta_3] \to (A, B \parallel_C D \to A \parallel_{B,C} D).$$

Recall that $\Delta_3$ is a basic diagram. Thus, by Definition 4, set $[\Delta_3]$ is empty. Therefore, $\vdash A, B \parallel_C D \to A \parallel_{B,C} D$. □

PROPOSITION 3.

$\vdash B \parallel_A C \wedge E \parallel_B D \wedge D \parallel_C F \wedge E \parallel_D F \wedge A \parallel_E F \to E \parallel_A F$.

PROOF. Consider diagram $\Delta_2$ depicted in Figure 3. As we have shown in Section 4, this diagram renders tuple

$$(\varnothing, \varnothing, E; \varnothing, F, \varnothing; A, \varnothing, \varnothing; \varnothing).$$

Thus, by the Diagram axiom,

$$\vdash [\Delta_2] \to (A \parallel_E F \to E \parallel_A F).$$

Recall from Section 4 that

$$[\Delta_2] = \{(B \parallel_A C), (E \parallel_B D), (D \parallel_C F), (E \parallel_D F)\}.$$

Therefore,

$\vdash B \parallel_A C \wedge E \parallel_B D \wedge D \parallel_C F \wedge E \parallel_D F \wedge A \parallel_E F \to E \parallel_A F$.
□

As our final example, we prove the Exchange axiom mentioned in the introduction. Although it is a property of non-conditional independence, it can be rephrased in the language of the conditional independence.

PROPOSITION 4.

$$\vdash A, B \parallel_\varnothing C \to (A \parallel_\varnothing B \to A \parallel_\varnothing B, C).$$

PROOF. Suppose that $A, B \parallel_\varnothing C$. Thus, $A \parallel_B C$ by Proposition 2. Therefore, by Proposition 1 and due to the assumption $A \parallel_\varnothing B$, we can conclude that $A \parallel_\varnothing B, C$. □

# 7. SOUNDNESS

We prove soundness of each axiom as a separate lemma.

LEMMA 1 (SYMMETRY). *For any protocol* $\mathcal{P} = (V, R)$, *if* $\mathcal{P} \vDash A \parallel_C B$, *then* $\mathcal{P} \vDash B \parallel_C A$.

PROOF. Assume that $r_1 \equiv_C r_2$ for some runs $r_1, r_2 \in R$. Thus, $r_2 \equiv_C r_1$. Hence, by the assumption of the lemma, there is $r \in R$ such that $r_2 \equiv_{A,C} r \equiv_{B,C} r_1$. Therefore, $r_1 \equiv_{B,C} r \equiv_{A,C} r_2$. □

LEMMA 2 (MONOTONICITY). *For any* $\mathcal{P} = (V, R)$, *if* $\mathcal{P} \vDash A \parallel_C B, D$, *then* $\mathcal{P} \vDash A \parallel_C B$.

PROOF. Assume that $r_1 \equiv_C r_2$ for some runs $r_1, r_2 \in R$. Hence, by the assumption of the lemma, there is $r \in R$ such that $r_1 \equiv_{A,C} r \equiv_{B,D,C} r_2$. Therefore, $r_1 \equiv_{A,C} r \equiv_{B,C} r_2$. □

Next, we establish a technical lemma that is used in the proof of soundness of the Diagram axiom.

LEMMA 3. *For any diagram* $\Delta \in Diag(Q)$ *and any protocol* $\mathcal{P} = (V, R)$ *such that* $\mathcal{P} \vDash \delta$ *for each* $\delta \in [\Delta]$, *if* $r^+, r^- \in R$ *and* $r^+ \equiv_Q r^-$, *then there is a function* $\rho$ *that maps vertices of the diagram* $\Delta$ *into runs in $R$ that satisfies the following conditions:*

1. $\rho(v_\Delta^+) = r^+$ *and* $\rho(v_\Delta^-) = r^-$,

2. *if* $v_1 \sim_S v_2$, *then* $\rho(v_1) \equiv_S \rho(v_2)$.

PROOF. Induction on the number of vertices in diagram $\Delta$. If $\Delta$ is a basic diagram, then define $\rho$ to be such that $\rho(v_\Delta^+) = r^+$ and $\rho(v_\Delta^-) = r^-$. Condition 2 is satisfied because of the assumption $r^+ \equiv_Q r^-$.

Suppose now that diagram $\Delta'$ is obtained from diagram $\Delta$ by adding a new vertex $v_{new}$, connected to vertices $u$ and $v$ by edges labeled with sets $A \cup C$ and $B \cup C$ respectively, such that $u \sim_C v$. By the induction hypothesis, there is a function $\rho$ on the vertices of the diagram $\Delta$ that satisfies conditions 1. and 2. of this lemma. In particular, $\rho(u) \equiv_C \rho(v)$. We will show how function $\rho$ could be extended to the vertex $v_{new}$ preserving conditions 1. and 2.

Note that $A \parallel_C B \in [\Delta']$, by Definition 4. Hence, by the assumption of this lemma, $\mathcal{P} \vDash A \parallel_C B$. Therefore, there is a run $r \in R$ such that $\rho(u) \equiv_{A,C} r \equiv_{B,C} \rho(v)$. Define $\rho(v_{new}) = r$.

To finish the proof of the lemma, we need to show that if $v_{new} \sim_S w$, where $w \neq v_{new}$ is a vertex in the diagram $\Delta'$, then $\rho(v_{new}) \equiv_S \rho(w)$. Note that vertex $w$ is also a vertex in the diagram $\Delta$, because $w \neq v_{new}$. Thus, Set $S$ could be partitioned into sets $S_1$ and $S_2$ such that: $S_1 \subset A \cup C$, $S_2 \subset B \cup C$, $u \sim_{S_1} w$ and $v \sim_{S_2} w$. Hence, by the induction hypothesis, $\rho(u) \equiv_{S_1} \rho(w)$ and $\rho(v) \equiv_{S_2} \rho(w)$. Thus,

$$\rho(v_{new}) = r \equiv_{S_1} \rho(u) \equiv_{S_1} \rho(w),$$

$$\rho(v_{new}) = r \equiv_{S_2} \rho(v) \equiv_{S_2} \rho(w).$$

Therefore, $\rho(v_{new}) \equiv_S \rho(w)$. □

LEMMA 4 (DIAGRAM). *For any protocol* $\mathcal{P} = (V, R)$, *if*

1. *diagram* $\Delta$ *renders tuple*

$$(A_1, A_2, A_3; B_1, B_2, B_3; C_1, C_2, C_3; D),$$

2. $\mathcal{P} \vDash \delta$ for each $\delta \in [\Delta]$,

3. $\mathcal{P} \vDash A_1, B_1, C_1 \parallel_{A_3,B_3,C_3,D} A_2, B_2, C_2$

then $\mathcal{P} \vDash A_1, A_2, A_3 \parallel_{C_1,C_2,C_3} B_1, B_2, B_3$.

PROOF. Let $r^+, r^- \in R$ be such that $r^+ \equiv_{C_1,C_2,C_3} r^-$. We will prove the existence of a run $r \in R$ such that

$$r^+ \equiv_{A_1,A_2,A_3,C_1,C_2,C_3} r,$$

$$r^- \equiv_{B_1,B_2,B_3,C_1,C_2,C_3} r.$$

By Lemma 3, there is a function $\rho$ that maps vertices of the diagram into runs of the protocol $\mathcal{P}$ that satisfies conditions 1. and 2. of Lemma 3.

By Definition 5, there are vertices $w_1$ and $w_2$ in the diagram $\Delta$ that satisfy conditions 1.-5. of that definition. In particular, $w_1 \sim_{A_3,C_3} v_\Delta^+$ and $w_2 \sim_{A_3,C_3} v_\Delta^+$. Thus, $w_1 \sim_{A_3,C_3} w_2$. Similarly, $w_1 \sim_{B_3,C_3} w_2$. By condition 5. of Definition 5, $w_1 \sim_D w_2$. Therefore, $w_1 \sim_{A_3,B_3,C_3,D} w_2$. Thus, by the assumption 3. of this lemma, there is a run $r \in R$ such that

$$\rho(w_1) \equiv_{A_1,B_1,C_1,A_3,B_3,C_3,D} r \qquad (1)$$

$$\rho(w_2) \equiv_{A_2,B_2,C_2,A_3,B_3,C_3,D} r \qquad (2)$$

By Definition 5,

$$w_1 \sim_{A_1,A_3,C_1,C_3} v_\Delta^+$$

$$w_2 \sim_{A_2,A_3,C_2,C_3} v_\Delta^+.$$

Hence, by condition 2. of Lemma 3,

$$\rho(w_1) \equiv_{A_1,A_3,C_1,C_3} r^+$$

$$\rho(w_2) \equiv_{A_2,A_3,C_2,C_3} r^+.$$

Finally, taking into account equations (1) and (2),

$$r^+ \equiv_{A_1,A_2,A_3,C_1,C_2,C_3} r.$$

Similarly, $r^- \equiv_{B_1,B_2,B_3,C_1,C_2,C_3} r$. □

## 8. COMPLETENESS

In the rest of the paper we establish completeness of our logical system.

THEOREM 2 (COMPLETENESS). *For any $\varphi \in \Phi$, if $\nvdash \varphi$, then there is a protocol $\mathcal{P}$ such that $\mathcal{P} \nvDash \varphi$.*

Suppose that $\nvdash \varphi$. Let $X$ be any maximal consistent subset of $\Phi$ containing formula $\neg\varphi$.

### 8.1 Chains of Diagrams

The chains of diagrams is a technical construction that we use to prove of the completeness theorem.

DEFINITION 6. *A Q-chain is an infinite sequence of diagrams $\Delta_0, \Delta_1, \ldots, \Delta_n, \ldots$ from $Diag(Q)$ such that $\Delta_0$ is the basic diagram and diagram $\Delta_{i+1}$ is an extension of the diagram $\Delta_i$ for each $i \geq 0$.*

LEMMA 5. *For any Q-chain $\Delta_0, \Delta_1, \ldots, \Delta_n, \ldots$, any label $p$, any $n$, and any $N \geq n$, if $x$ and $y$ are vertices on a diagram $\Delta_n$ and $x \sim_p y$ on diagram $\Delta_N$, then $x \sim_p y$ on diagram $\Delta_n$.*

PROOF. Suppose that there is $n \leq k < N$ such that $x \sim_p y$ on diagram $\Delta_{k+1}$, but not on diagram $\Delta_k$. By Definition 4, diagram $\Delta_{k+1}$ is obtained from diagram $\Delta_k$ by adding vertex $w$ connected to vertices $u$ and $v$ by edges labeled with sets $A \cup C$ and $B \cup C$, such that $u \sim_C v$ on diagram $\Delta_k$.

Since $x \sim_p y$ on diagram $\Delta_{k+1}$, but not on diagram $\Delta_k$, there must be a path labeled by $p$ between vertices $x$ and $y$ on diagram $\Delta_{k+1}$ that goes through both added edges: $(u,w)$ and $(w,v)$. Hence, $p \in (A \cup C) \cap (B \cup C)$. By Definition 1, sets $A$, $B$, and $C$ are disjoint. Thus, $p \in C$. Recall, however, that $u \sim_C v$ on diagram $\Delta_k$. Therefore, $x \sim_p y$ on diagram $\Delta_k$, which is a contradiction with the choice of $k$. □

DEFINITION 7. *A Q-chain $\Delta_0, \Delta_1, \Delta_2, \ldots$ is called sound if $[\Delta_n] \subseteq X$ for each $n \geq 0$.*

DEFINITION 8. *A Q-chain $\Delta_0, \Delta_1, \Delta_2, \ldots$ is complete if for any $A \parallel_C B \in X$, for any $n \geq 0$ and any two vertices $u, v$ of the diagram $\Delta_n$ such that $u \sim_C v$, there is $N \geq n$ and a vertex $w$ in the diagram $\Delta_N$ such that relations $u \sim_{A,C} w$ and $w \sim_{B,C} v$ hold in diagram $\Delta_N$.*

LEMMA 6. *For any set of secrets $Q$, there is a Q-chain which is complete and sound with respect to the set $X$.*

PROOF. The statement of the lemma follows from the Definition 4 and the fact that set $X$ is countable. □

### 8.2 Chain Protocol

We now show how a chain of diagrams can be converted into a protocol with certain desirable properties. Later, several such protocols will be combined into one in order to finish the proof of the completeness theorem.

LEMMA 7. *For each finite set of secrets $Q$ there is a protocol $\mathcal{P}$ such that*

1. *protocol $\mathcal{P}$ has at least one run,*

2. *$\mathcal{P} \vDash A \parallel_C B$ for each sets of secret variables $A$, $B$, and $C$ such that $A \parallel_C B \in X$,*

3. *$\mathcal{P} \nvDash P \parallel_Q R$ for each sets of secret variables $P$ and $R$ such that $P \parallel_Q R \notin X$.*

PROOF. By Lemma 6, there is Q-chain of diagrams

$$\Delta_0, \Delta_1, \Delta_2, \ldots,$$

which is complete and sound with respect to the set $X$. Let $V_0 \subset V_1 \subset V_2 \subset \ldots$ be the sets of vertices of these diagrams.

For any label $a$ and any two vertices $u, v \in \bigcup_i V_i$, we say that vertices $u$ and $v$ are $a$-equivalent if there is $k$ such that $u \sim_a v$ in diagram $\Delta_k$. Let $Val(a)$ be the set of equivalence classes on $\bigcup_i V_i$ with respect to this equivalence relation.

For any $v \in \bigcup_i V_i$ and any label $a$, define function $r_v(a)$ to be equal to the $a$-equivalence class of $v$:

$$r_v(a) = [v]_a.$$

Let $\mathcal{R} = \{r_v \mid v \in \bigcup_i V_i\}$. This concludes the definition of the protocol $\mathcal{P} = (Val, \mathcal{R})$. We will now show that this protocol satisfies conditions 1., 2., and 3. of the lemma.

To prove the first condition, notice that set $\bigcup_i V_i$ is not empty, because it contains vertices $v^+$ and $v^-$ from the basic diagram $\Delta_0$. Thus, set $\{r_v \mid v \in \bigcup_i V_i\}$ is also not empty.

To prove the second condition, consider any $r_u, r_v \in \mathcal{R}$ such that $r_u \equiv_C r_v$. We will show that there is $r_w \in \mathcal{R}$ such that $r_u \equiv_{A,C} r_w \equiv_{B,C} r_v$. Indeed, $r_u \equiv_C r_v$ implies that $[u]_c = [v]_c$ for each $c \in C$. Thus, for each $c \in C$, vertices $u$ and $v$ are $c$-equivalent. Hence, there must exists $n \geq 0$ such that $u \sim_C v$ in $\Delta_n$. By Definition 8, there is $N \geq n$ and a vertex $w$ in the diagram $\Delta_N$ such that relations $u \sim_{A,C} w$ and $w \sim_{B,C} v$ hold in diagram $\Delta_N$. Thus, $[u]_x = [w]_x$ for each $x \in A \cup C$ and $[w]_y = [v]_y$ for each $y \in B \cup C$. Therefore, $r_u \equiv_{A,C} r_w \equiv_{B,C} r_v$.

To prove the third condition, assume the opposite: $\mathcal{P} \vDash P \parallel_Q R$. Consider vertices $v^+$ and $v^-$ of the based diagram $\Delta_0$. By Definition 4, $v^+ \sim_Q v^-$ on diagram $\Delta_0$. Thus, $[v^+]_q = [v^-]_q$ for each $q \in Q$. Hence, $r_{v^+} \equiv_Q r_{v^-}$. Then, by the assumption $\mathcal{P} \vDash P \parallel_Q R$, there must be a run $r_w$ such that $r_{v^+} \equiv_{P,Q} r_w \equiv_{R,Q} r_{v^-}$. Hence, $[v^+]_t = [w]_t$ for each $t \in P \cup Q$ and $[w]_t = [v^-]_t$ for each $t \in R \cup Q$. Thus,

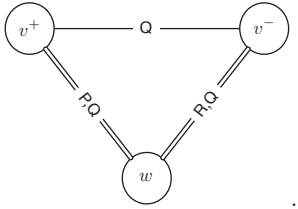

.

Let $n$ be the smallest integer such that $\Delta_n$ contains vertex $w$. By Definition 4, there are vertices $u$ and $v$ in diagram $\Delta_n$ such that

1. vertex $w$ is only connected in diagram $\Delta_n$ to $u$ and $v$,
2. edge $(u, w)$ is labeled with a set $A$,
3. edge $(w, v)$ is labeled with set $B$,
4. $u \sim_{A \cap B} v$ in diagram $\Delta_{n-1}$, and
5. $A \setminus B \parallel_{A \cap B} B \setminus A \in [\Delta_n]$.

Since chain $\Delta_0, \Delta_1, \ldots$ is sound with respect to set $X$, the last condition above implies that

$$A \setminus B \parallel_{A \cap B} B \setminus A \in X.$$

By Monotonicity axiom,

$$X \vdash A \setminus B \parallel_{A \cap B} P \cap (B \setminus A), R \cap (B \setminus A), Q \cap (B \setminus A).$$

By Symmetry axiom,

$$X \vdash P \cap (B \setminus A), R \cap (B \setminus A), Q \cap (B \setminus A) \parallel_{A \cap B} A \setminus B.$$

By Monotonicity axiom,

$$X \vdash\ P \cap (B \setminus A), R \cap (B \setminus A), Q \cap (B \setminus A) \parallel_{A \cap B}$$
$$P \cap (A \setminus B), R \cap (A \setminus B), Q \cap (A \setminus B).$$

Again by Symmetry axiom,

$$X \vdash\ P \cap (A \setminus B), R \cap (A \setminus B), Q \cap (A \setminus B) \parallel_{A \cap B}$$
$$P \cap (B \setminus A), R \cap (B \setminus A), Q \cap (B \setminus A).$$

In other words,

$$X \vdash\ P \cap (A \setminus B), R \cap (A \setminus B), Q \cap (A \setminus B)$$
$$\parallel_{P \cap (A \cap B), R \cap (A \cap B), Q \cap (A \cap B), (A \cap B) \setminus (P \cup Q \cup R)}$$
$$P \cap (B \setminus A), R \cap (B \setminus A), Q \cap (B \setminus A).$$

We now apply the Diagram axiom (see Figure 4) with

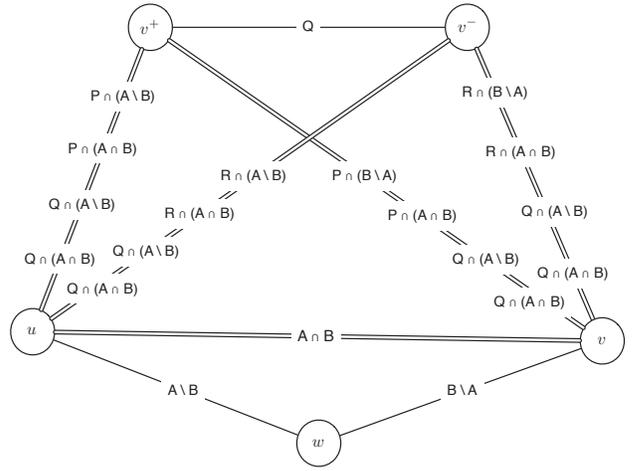

**Figure 4: Diagram $\Delta_n$.**

$$A_1 = P \cap (A \setminus B) \qquad A_2 = P \cap (B \setminus A)$$

$$B_1 = R \cap (A \setminus B) \qquad B_2 = R \cap (B \setminus A)$$

$$C_1 = Q \cap (A \setminus B) \qquad C_2 = Q \cap (B \setminus A)$$

$$A_3 = P \cap (A \cap B)$$

$$B_3 = R \cap (A \cap B)$$

$$C_3 = Q \cap (A \cap B)$$

$$D = (A \cap B) \setminus (P \cup Q \cup R)$$

to conclude that

$$X \vdash\ P \cap (A \setminus B), P \cap (B \setminus A), P \cap (A \cap B)$$
$$\parallel_{Q \cap (A \setminus B), Q \cap (B \setminus A), Q \cap (A \cap B)} \qquad (3)$$
$$R \cap (A \setminus B), R \cap (B \setminus A), R \cap (A \cap B).$$

Recall that $w \sim_{P \cup Q} v^+$ and $w \sim_{R \cup Q} v^-$. At the same time, vertex $w$ is only connected in diagram $\Delta_n$ to $u$ and $v$, edge $(u, w)$ is labeled with a set $A$, and edge $(w, v)$ is labeled with set $B$. Hence, $P \cup Q \subseteq A \cup B$. Thus, statement (3) implies that $X \vdash P \parallel_Q R$, which is a contradiction with the assumption. $\square$

## 8.3 Protocol Composition

In this section we introduce a way to combine several different protocols over $(S, G)$ into a single protocol.

DEFINITION 9. *For any protocols $\mathcal{P}_1 = (V_1, R_1), \ldots, \mathcal{P}_n = (V_n, R_n)$, let $\mathcal{P}_1 \times \cdots \times \mathcal{P}_n$ be a protocol $(V, R)$ such that*

1. $V(a) = V_1(a) \times \cdots \times V_n(a)$, *for each $a \in S$,*
2. $R$ *is a set of all functions $r(x) = \langle r_1(x), \ldots, r_n(x) \rangle$ for all $r_1 \in R_1, \ldots, r_n \in R_n$.*

LEMMA 8. *Let $\mathcal{P}_1 = (V_1, R_1), \ldots, \mathcal{P}_n = (V_n, R_n)$ be protocols such that set $R_k$ is not empty for each $k \leq n$. Then $\mathcal{P}_1 \times \cdots \times \mathcal{P}_n \vDash A \parallel_C B$ if and only if $\mathcal{P}_k \vDash A \parallel_C B$ for each $k \leq n$.*

PROOF. ($\Rightarrow$) : Suppose that $r_k^1, r_k^2 \in R_k$ are such that $r_k^1 \equiv_C r_k^2$. We will show that there is a run $r_k \in R_k$ such that $r_k^1 \equiv_{A,C} r_k \equiv_{B,C} r_k^2$.

Let $(V, R)$ be protocol $\mathcal{P}_1 \times \cdots \times \mathcal{P}_n$. Consider any runs

$$r_1 \in R_1, \ldots, r_{k-1} \in R_{k-1}, r_{k+1} \in R_{k+1}, \ldots, r_n \in R_n.$$

Such runs exists due to the assumption of the lemma. Let $r^1, r^2 \in R$ be such that for each secret variable $x$,

$$r^1(x) = \langle r_1(x), \ldots, r_{k-1}(x), r_k^1(x), r_{k+1}(x), \ldots, r_n(x)\rangle,$$

$$r^2(x) = \langle r_1(x), \ldots, r_{k-1}(x), r_k^2(x), r_{k+1}(x), \ldots, r_n(x)\rangle.$$

Note that $r_k^1 \equiv_C r_k^2$ implies that $r^1(c) \equiv_C r^2(c)$. Hence, by the assumption of the lemma, there is a run $r \in R$ such that $r^1 \equiv_{A,C} r \equiv_{B,C} r^2$. Let $r_k(x)$ be defined to be the $k$-th component of $r(x)$ for each secret variable $x$. Thus, by Definition 9, $r_k \in R_k$. Finally, $r^1 \equiv_{A,C} r \equiv_{B,C} r^2$ implies that $r_k^1 \equiv_{A,C} r_k \equiv_{B,C} r_k^2$.

($\Leftarrow$) : Suppose that $r^1, r^2 \in R$ are such that $r^1 \equiv_C r^2$. We will show that there is $r \in R$ such that $r^1 \equiv_{A,C} r \equiv_{B,C} r^2$. Assume that $r^1(x) = \langle r_1^1(x), \ldots, r_n^1(x)\rangle$, and $r^2(x) = \langle r_1^2(x), \ldots, r_n^2(x)\rangle$. Assumption $r^1 \equiv_C r^2$ implies that $r_k^1 \equiv_C r_k^2$ for each $k \le n$. Thus, by the assumption of the lemma, there are runs $r_1 \in R_1, \ldots, r_n \in R_n$ such that $r_k^1 \equiv_{A,C} r_k \equiv_{B,C} r_k^2$ for each $k \le n$. Define $r(x) = \langle r_1(x), \ldots, r_n(x)\rangle$, Therefore, $r^1 \equiv_{A,C} r \equiv_{B,C} r^2$. □

### 8.4 Completeness: final steps

We are now ready to finish the proof of the completeness theorem. Let $S$ be the finite set of all variables that appear in the formula $\varphi$. Let $Q_1, \ldots, Q_n$ be all subsets of $S$. By Lemma 7, there are protocols $\mathcal{P}_1, \ldots, \mathcal{P}_n$ such that

1. $\mathcal{P}_k \vDash A \parallel_C B$ for each sets of secret variables $A$, $B$, and $C$ such that $A \parallel_C B \in X$,

2. $\mathcal{P}_k \nvDash P \parallel_{Q_k} R$ for each sets of secret variables $P$ and $R$ such that $P \parallel_{Q_k} R \notin X$.

Let $\mathcal{P} = \mathcal{P}_1 \times \cdots \times \mathcal{P}_n$.

LEMMA 9. *For each $\psi \in \Phi$ that only uses secret variables from set $S$, $\mathcal{P} \vDash \psi$ if and only if $\psi \in X$.*

PROOF. Induction on the structural complexity of formula $\psi$. Case $\psi$ being $\bot$ follows from the assumption of consistency of $X$ and Definition 3. The induction case $\psi \equiv \psi_1 \to \psi_2$ follows from the maximality and consistence of set $X$ in the standard way. We are only left to consider the case when $\psi$ is an atomic formula $P \parallel_Q R$ for some $P, Q, R \subseteq S$. Assume that $Q = Q_{k_0}$.
($\Rightarrow$) : Suppose that $X \nvdash P \parallel_Q R$. Thus, $\mathcal{P}_{k_0} \nvDash P \parallel_Q R$ due to the choice of the protocol $\mathcal{P}_{k_0}$. Note that each of the protocols $\mathcal{P}_1, \ldots, \mathcal{P}_n$ has at least one run due to Lemma 7. Thus, by Lemma 8, $\mathcal{P} \nvDash P \parallel_Q R$.
($\Leftarrow$) : If $X \vdash P \parallel_Q R$, then, $\mathcal{P}_k \nvDash P \parallel_Q R$ for each $k \le n$ due to the choice of the protocols $\mathcal{P}_1, \ldots, \mathcal{P}_n$. Note again that each of the protocols $\mathcal{P}_1, \ldots, \mathcal{P}_n$ has at least one run due to Lemma 7. Thus, by Lemma 8, $\mathcal{P} \vDash P \parallel_Q R$. □

Recall now that $\neg \varphi \in X$. Hence, $\varphi \notin X$ due to consistency of $X$. Therefore, $\mathcal{P} \nvDash \varphi$ by Lemma 9. This concludes the proof of Theorem 2. □

## 9. CONCLUSION

In this paper we gave a recursively enumerable axiomatization of propositional properties of relation $A \parallel_C B$, assuming that sets $A$, $B$, and $C$ are pair-wise disjoint. Although Definition 3 is meaningful if the sets are not disjoint, our completeness proof will not work (see Lemma 5). At the same time, it is interesting to point out that due to Definition 3, statement $B \parallel_A B$ means that *any two runs that agree on $A$ also agree on $B$*. Thus, $B \parallel_A B$ represents *functional dependency* relation between values of $A$ and $B$. Functional dependency alone was axiomatized by Armstrong [1]. It appears that allowing sets $A$, $B$, and $C$ to be non-disjoint leads to a significantly more powerful language. Complete axiomatization of all properties expressible in such language remains an open question.

## 10. REFERENCES

[1] W. W. Armstrong. Dependency structures of data base relationships. In *Information processing 74 (Proc. IFIP Congress, Stockholm, 1974)*, pages 580–583. North-Holland, Amsterdam, 1974.

[2] Michael S. Donders, Sara Miner More, and Pavel Naumov. Information flow on directed acyclic graphs. In Lev D. Beklemishev and Ruy de Queiroz, editors, *WoLLIC*, volume 6642 of *Lecture Notes in Computer Science*, pages 95–109. Springer, 2011.

[3] Dan Geiger, Azaria Paz, and Judea Pearl. Axioms and algorithms for inferences involving probabilistic independence. *Inform. and Comput.*, 91(1):128–141, 1991.

[4] Erich Grädel and Jouko Väänänen. Dependence and Independence. To appear in Studia Logica.

[5] Erich Grädel and Jouko Väänänen. Dependence, Independence, and Incomplete Information. In *Proceedings of 15th International Conference on Database Theory, ICDT 2012*, 2012.

[6] Joseph Y. Halpern and Kevin R. O'Neill. Secrecy in multiagent systems. *ACM Trans. Inf. Syst. Secur.*, 12(1):1–47, 2008.

[7] Christian Herrmann. On the undecidability of implications between embedded multivalued database dependencies. *Inf. Comput.*, 122(2):221–235, 1995.

[8] Christian Herrmann. Corrigendum to "on the undecidability of implications between embedded multivalued database dependencies" [inform. and comput. 122(1995) 221-235]. *Inf. Comput.*, 204(12):1847–1851, 2006.

[9] Jérôme Lang, Paolo Liberatore, and Pierre Marquis. Conditional independence in propositional logic. *Artif. Intell.*, 141(1/2):79–121, 2002.

[10] Sara Miner More and Pavel Naumov. An independence relation for sets of secrets. In H. Ono, M. Kanazawa, and R. de Queiroz, editors, *Proceedings of 16th Workshop on Logic, Language, Information and Computation (Tokyo, 2009), LNAI 5514*, pages 296–304. Springer, 2009.

[11] Sara Miner More and Pavel Naumov. Hypergraphs of multiparty secrets. In *11th International Workshop on Computational Logic in Multi-Agent Systems CLIMA XI (Lisbon, Portugal), LNAI 6245*, pages 15–32. Springer, 2010.

[12] Sara Miner More and Pavel Naumov. Logic of secrets in collaboration networks. *Ann. Pure Appl. Logic*, 162(12):959–969, 2011.

[13] Sara Miner More, Pavel Naumov, Brittany Nicholls, and Andrew Yang. A ternary knowledge relation on secrets. In Krzysztof R. Apt, editor, *Proceedings of the 13th Conference on Theoretical Aspects of Rationality and Knowledge (TARK-2011), Groningen, The Netherlands, July 12-14, 2011*, pages 46–54. ACM, 2011.

[14] Sara Miner More, Pavel Naumov, and Benjamin Sapp. Concurrency semantics for the Geiger-Paz-Pearl axioms of independence. In Marc Bezem, editor, *20th Annual Conference on Computer Science Logic, , CSL 2011, September 12-15, 2011, Bergen, Norway, Proceedings*, volume 12 of *LIPIcs*, pages 443–457. Schloss Dagstuhl - Leibniz-Zentrum fuer Informatik, 2011.

[15] Pavel Naumov and Brittany Nicholls. Game semantics for the Geiger-Paz-Pearl axioms of independence. In *The Third International Workshop on Logic, Rationality and Interaction (LORI-III), LNAI 6953*, pages 220–232. Springer, 2011.

[16] D. Stott Parker, Jr. and Kamran Parsaye-Ghomi. Inferences involving embedded multivalued dependencies and transitive dependencies. In *Proceedings of the 1980 ACM SIGMOD international conference on Management of data*, SIGMOD '80, pages 52–57, New York, NY, USA, 1980. ACM.

[17] Milan Studený. Conditional independence relations have no finite complete characterization. In *Information Theory, Statistical Decision Functions and Random Processes. Transactions of the 11th Prague Conference vol. B*, pages 377–396. Kluwer, 1990.

[18] David Sutherland. A model of information. In *Proceedings of Ninth National Computer Security Conference*, pages 175–183, 1986.